\documentclass[conference,letterpaper,10pt]{IEEEtran}
\usepackage[top=0.75in,bottom=1.05in,left=0.625in,right=0.63in]{geometry}
\usepackage{rotating}
\usepackage{epstopdf}
\usepackage{cite}
\usepackage[utf8]{inputenc} 
\usepackage[T1]{fontenc}
\usepackage{url}
\usepackage{titlesec}
\titlespacing*{\subsection}{0pt}{0pt}{0pt}

\usepackage{ifthen}
\usepackage[switch]{lineno}
\usepackage{cite}
\usepackage[cmex10]{amsmath} 
\usepackage{amssymb}
\usepackage{graphicx}
\usepackage{helvet}
\usepackage{titlesec}
\usepackage{titlesec}
\titlespacing*{\subsection}{0pt}{*0.5}{*0.5}   

\usepackage{titlesec}
\titlespacing*{\section}{0pt}{*1}{*0.7} 

\usepackage{epstopdf}
\usepackage{epsfig}
\usepackage{titlesec}
\usepackage{multirow}
\usepackage{amsthm}
\usepackage{setspace}

\usepackage{comment}
\usepackage{caption}
\usepackage{amssymb}
\usepackage[cmex10]{amsmath}
\usepackage{breqn}
\usepackage{graphicx}
\usepackage{stfloats,adjustbox}
\usepackage{url}
\usepackage{xcolor}
\usepackage{multirow}
\usepackage{bm}
\usepackage{bbm}
\usepackage{amsmath}
\usepackage{enumitem}
\usepackage{subcaption}
\usepackage[english]{babel}
\usepackage{ragged2e}
\usepackage{blindtext}
\usepackage{caption, multirow, makecell}

\usepackage{fixltx2e}
\usepackage{longtable,eqnarray,makecell}
\usepackage{amsmath,amssymb}
\usepackage{threeparttable,stackengine}
\usepackage{tikz}
\usepackage{amsthm}
\newtheorem{example}{Example}
\newtheorem{remark}{Remark}

\usepackage{booktabs}

\newcounter{myitemcounter}
\begin{document}
\title{Hierarchical Coded Caching in High Memory Regime with Coded Placement} 

\author{\IEEEauthorblockN{Rajlaxmi Pandey}
\IEEEauthorblockA{\textit{Dept. of Electrical Communication } \\
\textit{Engineering, Indian Institute of Science, } \\
Bengaluru, India \\
Email: rajlaxmip@iisc.ac.in}
\and
\IEEEauthorblockN{Charul Rajput}
\IEEEauthorblockA{\textit{Dept. of Mathematics and Systems } \\
\textit{Analysis, Aalto University,}\\
Finland \\
Email: charul.rajput@aalto.fi}
\and
\IEEEauthorblockN{B. Sundar Rajan}
\IEEEauthorblockA{\textit{Dept. of Electrical Communication } \\
\textit{Engineering, Indian Institute of Science, } \\
Bengaluru, India \\
Email: bsrajan@iisc.ac.in}
}

\maketitle

\begin{abstract}

We consider a two-layer hierarchical coded caching network where a server with a library of \( N \) files is connected to \( K_1 \) mirrors, each having a cache memory of size \( M_1 \). Each mirror is further connected to \( K_2 \) users, each equipped with a dedicated cache of size \( M_2 \). In this paper, we propose two distinct coded caching schemes based on coded placement, corresponding to two distinct memory pairs, \( (M_1, M_2) \), which operate effectively in the high memory regime, i.e., when both \( M_1 \) and \( M_2 \) are large. We show that the proposed schemes outperform existing schemes at these memory points for smaller values of \( K_2 \). In setups where mirrors are positioned near each other, avoiding signal interference is crucial. This can be ensured by having all mirrors transmit using orthogonal carrier frequencies. To compare our schemes with existing ones, we used the composite rate metric, which accurately represents the total bandwidth utilized in such setups. The composite rate is given by \( \overline{R} = R_1 + K_1 R_2 \), where \( R_1 \) is the rate from the server to the mirrors, and \( R_2 \) is the rate from the mirrors to the users, with respect to \( M_1 \) and \( M_2 \).

\end{abstract}
\begin{IEEEkeywords}
 Hierarchical coded caching, Transmission rate, Coded Placement.
\end{IEEEkeywords}
\IEEEpeerreviewmaketitle 


\IEEEpeerreviewmaketitle
\section{Introduction}
\label{intro}

In the last decade, the exponential growth in data consumption has been largely fueled by the proliferation of smartphones and other electronic devices. To address the challenges posed by increased data traffic, particularly during peak usage periods, coded caching has emerged as an effective strategy. The foundational study by \cite{maddah2014fundamental} introduced a framework that combines cache placement and delivery phases, enabling efficient cache management and data transmission. In the cache placement phase, which occurs during off-peak times, the central server pre-stores content for users. The delivery phase is dedicated to transmitting files that are not already cached. The term ``transmission rate'' refers to the total number of bits that the server must transmit to fulfill the requests of all users.

In a coded caching system, the network consists of one server with access to a library of \( N \) files, all of equal length, connected to \( K \) users via a shared error-free link. Each user has the capability to cache up to \( M \) files. For this setup, the scheme proposed in \cite{maddah2014fundamental}, henceforth referred to as the MN scheme, is demonstrated to be optimal for scenarios with distinct demands when employing uncoded prefetching \cite{YMA}. In the basic coded caching scheme, each file hosted on the server must be divided into a significant number of subpackets, defined by the subpacketization level. In \cite{yan2017placement}, the authors proposed a scheme utilizing a placement delivery array (PDA) to characterize both the placement and delivery phases simultaneously. They demonstrated that the MN scheme can be represented using a specific type of PDA, referred to as the MN PDA.


In real-world scenarios, caching systems consist of multiple layers connecting the central server to end-users, with intermediate devices functioning as mirrors to enhance connectivity. The work in \cite{karamchandani2016hierarchical,8481555,wang2019reduce,kong2023combinatorial} investigates a model where the server communicates through an error-free broadcast link to $K_{1}$ mirror sites. Each of these mirrors subsequently links to $K_{2}$ users, also via error-free broadcast links. This results in a total user count $K$ given by $K = K_{1}K_{2}$. Let $R_{1}$ denote the transmission rate (normalized by file size) from the server to the mirrors, and $R_{2}$ represent the rate from the mirrors to the users, both essential for fulfilling user demands.

\subsection{System Model}
\label{systemmodel}

\begin{figure}[t]
	\begin{center}
		\captionsetup{justification = centering}
		\includegraphics[width = 0.75\columnwidth]{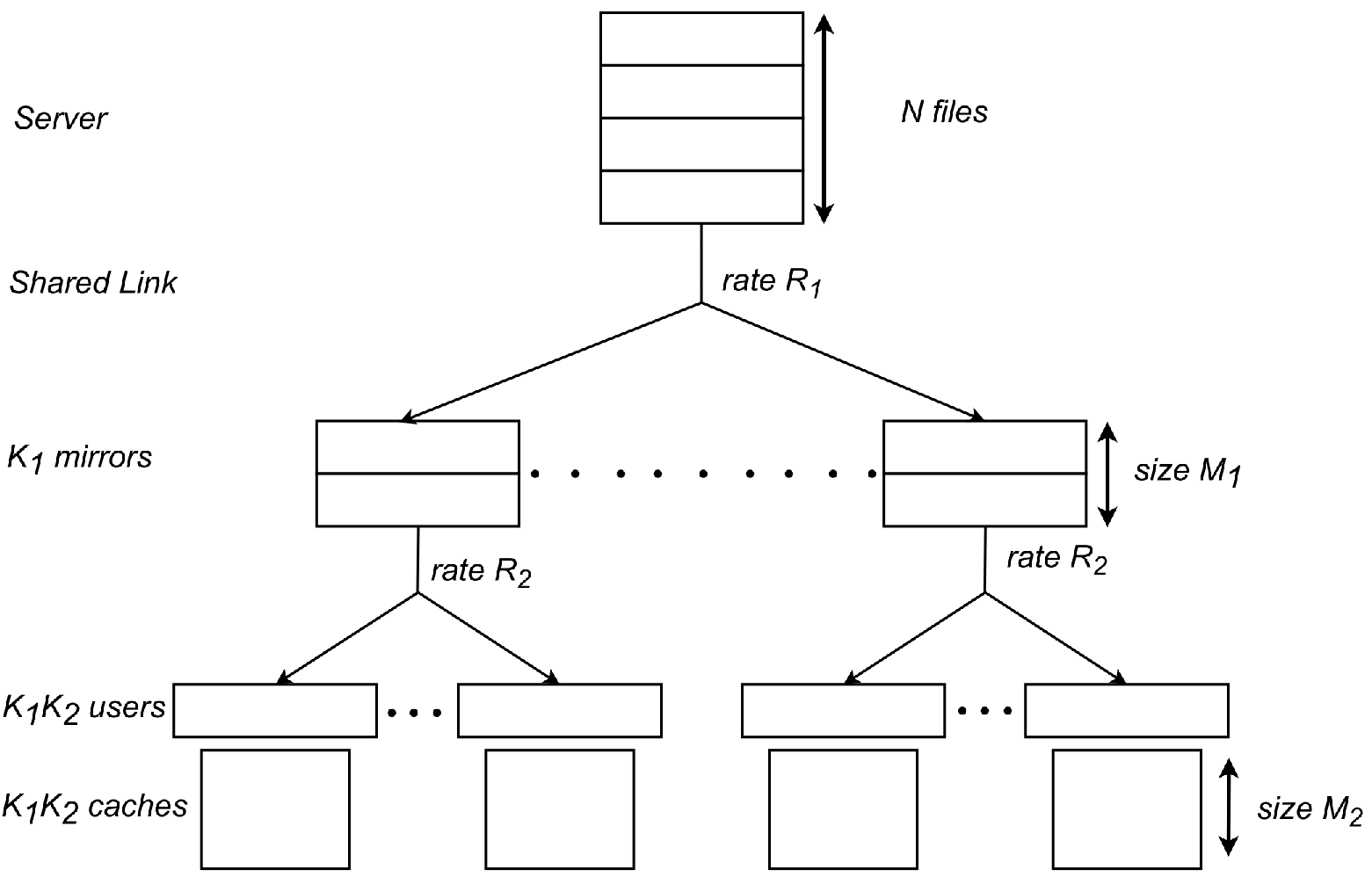}
		\caption{Hierarchical two-layer network.}
		\label{fig:setting}
	\end{center}
\end{figure}

A $(K_1, K_2; M_1, M_2; N)$ hierarchical coded caching system is illustrated in Fig. \ref{fig:setting}. The server stores a collection of $N$ files, each of size $F$ bits, denoted as $W_1, W_2, \dots, W_N$. There are $K_1$ mirrors, each connected to $K_2$ users via an error-free shared link. Every mirror and user is equipped with a cache of size $M_1$ and $M_2$ files, respectively. The system operates through the following three phases:

\begin{enumerate}[label=(\alph*)]
\item \emph{Placement phase}: During this phase, the server populates the cache of each mirror and user with parts of the file contents without knowing the specific demands of the users.
\item \emph{Server delivery phase}: Once the users reveal their demands, the server broadcasts either coded or uncoded multicast messages to the mirrors, with a total size of at most $R_1F$ bits per mirror.
\item \emph{Mirror delivery phase}: Each mirror, based on the server's transmissions and its own cache contents, generates and transmits coded or uncoded messages of size at most $R_2F$ bits to the users connected to it.
\end{enumerate}

In environments where mirrors are located in close proximity, it is essential to prevent signal interference. This can be accomplished by ensuring that all mirrors transmit using orthogonal carrier frequencies. This highlights the importance of the composite rate, which represents the total bandwidth utilized and is defined as:
\vspace{0.2cm}
\[
\overline{R} = R_1 + K_1 R_2.
\]

 \subsection{Notations}
For a positive integer \( n \), the notation \([n]\) represents the set \( \{1, 2, \ldots, n\} \). For two sets \( A \) and \( B \), \( A \setminus B \) denotes the set of elements that are in \( A \) but not in \( B \). The size of a set \( A \) is indicated by \( |A| \). 
The binomial coefficient \( \binom{n}{k} \) is defined as \( \frac{n!}{k!(n-k)!} \), where \( n \) and \( k \) are positive integers satisfying \( k \leq n \). For any negative integer \( k \), the value of \( \binom{n}{k} \) is defined to be zero. 

\subsection{Related Works}
Hierarchical coded caching was initially studied in \cite{karamchandani2016hierarchical}, where we refer to the proposed approach as the KNMD (Karamchandani, Niesen, Maddah-Ali, Diggavi) scheme. In this scheme, the server does not consider the cached content of the users when transmitting to the mirrors, potentially leading to redundant data transmissions at the first layer. In \cite{8481555}, referred to as the ZWXWLL (Zhang, Wang, Xiao, Wu, Liang, Li) scheme, the authors developed a hybrid caching scheme to reduce the traffic loads and transmission rates from both the server and the mirrors. They analytically derived the transmission rates and demonstrated that the hybrid caching scheme is capable of reducing the transmission rate from the server without increasing the transmission rate from each mirror compared to the caching scheme in \cite{karamchandani2016hierarchical}.
 In \cite{wang2019reduce}, the authors introduced the WWCY (Wang, Wu, Chen, Yin) caching scheme, which effectively leverages idle time resources by enabling simultaneous transmissions between the server and mirrors. Remarkably, the WWCY scheme achieves the minimum load \( R_2 \) when utilizing uncoded placement, and the load of the first layer, \( R_1 \), is reduced compared to the scheme presented in \cite{8481555}.
In \cite{liu2021intension}, the authors introduced the LZX (Liu, Zhang, Xie) scheme, focusing on a scenario involving a single mirror with two attached users. This scheme highlights the tension between the rates of the two layers. The authors derived new converse bounds and proposed novel achievable schemes that meet these bounds.
In \cite{kong2023combinatorial}, referred to as the KWC (Kong, Wu, Cheng) scheme, the authors proposed a hierarchical placement delivery array (HPDA), a combinatorial structure inspired by PDAs, to characterize both the placement and delivery phases of hierarchical coded caching. By combining any two PDAs, this approach enables flexible subpacketization. When MN PDAs are used as the base, the HPDA-based scheme reduces to the WWCY scheme from \cite{wang2019reduce}.

\subsection{Our Contributions}
The proposed scheme is inspired by the work in \cite{9745952}, which introduced a coded placement-based approach for the case \( N \leq K \leq 2N - 1 \) in a single-layer coded caching network. Building on this, our contributions are summarized as follows:

\begin{itemize}
    \item We propose two coded caching schemes for hierarchical two-layer networks, which show improved performance over state-of-the-art solutions for specific memory pairs \( M_1 \) and \( M_2 \), particularly when cache sizes are large for both mirrors and users.

    \item We provide a detailed comparison of our schemes with existing methods by plotting the composite rate $\overline{R}$ with respect to $M_1$ and $M_2$, illustrating improvements in both dimensions.

\end{itemize}

\section{Proposed Coded Caching Schemes for Hierarchical Networks} 

Building on the contributions outlined above, we now propose two coded caching schemes for hierarchical networks. Both schemes utilize coded placement to minimize transmission rates while ensuring efficient file delivery to users through mirrors. The schemes differ in their placement and delivery strategies.
\subsection{The First Scheme} \label{firsts}
In this scheme, we propose a coded caching scheme for the case \( N \leq K_1 K_2 \), operating for the memory pair \( M_1, M_2 \) as follows.

\begin{equation}
M_1 = \frac{(K-K_2)(K-K_2-1)N + K_2(K-2)N + K_2}{K(K-1)},
\label{eq:M1}
\end{equation}

\begin{equation}
M_2 = \frac{[(K-1)(K-2) - (K-K_2)(K-K_2-1)]N}{K(K-1)}.
\label{eq:M2}
\end{equation}

\subsubsection{Placement phase}
Each file \( W_n \) of size \( F \) bits is divided into \( 2 \binom{K}{2} \) subfiles, where each subfile is of size \( \frac{F}{2 \binom{K}{2}} \) bits, i.e., \( W_n \rightarrow W^{ij}_n \) for all \( n \in [N] \) and for all \( i,j \in [K] \) with \( i \neq j \). Let the cache content of mirror $m$ be denoted by $B_m$ for $m \in [K_1]$, and the cache content of user $k$ be denoted by $C_k$ for $k \in [K]$.
We use $U_m = \{ (m-1)K_2 + i \mid i = 1, 2, \ldots, K_2 \}$ to define the set of users which are connected to mirror $m$. 

The cache placement in the \( m \)-th mirror, denoted as \( B_m \), consists of the following three types of subfiles.

\begin{itemize}
\item  
\begin{align}
\{ W^{ij}_n \ | \ i, j \in [K] \setminus U_m, i \neq j, \forall n \in [N] \}.
\label{cachmirr1}
\end{align}

\item For all \( k \in U_m \),
\begin{align}
W^{k(k+1)}_n - W^{kj}_n, \quad \forall n \in [N], \quad j \in S_k,
\label{mp2}
\end{align}
where \( S_k = [K] \setminus \{k\} \). 

\item 
\begin{align}
\sum_{n=1}^{N} W^{k(k+1)}_n, \quad \forall k \in U_m.
\label{mp3}
\end{align}

The second and third types of placement compute specific functions of subfiles and store them in the cache of the \( m \)-th mirror. 

For the second type, given in \eqref{mp2}, the mirror caches the difference \( W^{k(k+1)}_n - W^{kj}_n \) for each file \( n \in [N] \), all \( k \in U_m \), and for each \( j \in S_k \), where \( S_k = [K] \setminus \{k\} \). This ensures that the cache stores linear combinations of subfiles related to the users connected to the mirror.

For the third type, given in \eqref{mp3}, the mirror caches the sum of all subfiles \( W^{k(k+1)}_n \) over all files \( n \in [N] \), for each \( k \in U_m \). This placement enables efficient encoding and decoding during the delivery phase, as the cached content includes a global function of the relevant subfiles.

\end{itemize}

The cache placement for user \( k \) is defined as follows
\begin{equation}
    C_k = \{ W^{ij}_n : \{i, j\} \cap U_m \neq \emptyset, \, i, j \in S_k, \, i \neq j, \, \forall n \in [N] \}.
    \label{userplace}
\end{equation}

 \subsubsection{Delivery phase}

In this section, we describe the entire delivery process, which is divided into two main stages:

1. \textbf{Server to Mirrors:} The server sends the required data to the mirrors. This stage involves transmitting both coded and uncoded subfiles, ensuring that each mirror receives the required subfiles to deliver to its associated users.
   
2. \textbf{Mirror to Users:} After receiving the data from the server, the mirrors begin transmitting the requested files to the users. This stage is further subdivided into five phases, labeled Transmission A through Transmission E. Each phase is designed to address a specific aspect of the file delivery process, ensuring efficient delivery by leveraging both coded and uncoded placement strategies.

Let the demand vector be $\overline{d} = (d_1, d_2, \ldots, d_K)$, where each file is requested by at least one user. Let $N^s_k$ represent the number of users in the set $S_k$ requesting file $W_{d_s}$.

\textbf{Server to Mirrors:}

For each user \( k \in [K] \), the server sends the following transmission to the mirrors:

\[
Y^k_{\overline{d}} = \sum_{s \in S_k} \left( \frac{\alpha^s_k}{N^s_k} \right) W^{ks}_{d_s},
\]
\begin{equation}
\label{SM}
\end{equation}
where:
\begin{itemize}
    \item \( d_s \) denotes the file requested by user \( s \),
    \item \( N^s_k \) is the number of users in the set \( S_k \) requesting file \( W_{d_s} \),
    \item \( \alpha^s_k = 1 - 2I(W_{d_k} = W_{d_s}) \), with \( I \) being the indicator function defined as:
\end{itemize}

\[
I(A) = 
\begin{cases}
1 & \text{if } A \text{ is true}, \\
0 & \text{if } A \text{ is false}.
\end{cases}
\]

\textbf{Mirror to Users:} The transmissions from the mirror \( m \) to users, \( m \in [K_1] \), are described as follows:

\textbf{Transmission A:} The mirror transmits all \( K \) coded subfiles received from the server as given in \eqref{SM}, after removing the subfiles available in the mirror's cache. For all \( k \in U_m \), the mirror sends
\begin{align}
     \sum_{s \in S_k} \left( \frac{\alpha^s_k}{N^s_k} \right) W^{ks}_{d_s}.
    \label{StepA1}
\end{align}

For all \( k \notin U_m \), the mirror sends

\begin{align}
    \sum_{s \in U_m} \left( \frac{\alpha^s_k}{N^s_k} \right) W^{ks}_{d_s}.
    \label{StepA2}
\end{align}

\textbf{Transmission B:} The mirror transmits the uncoded subfiles stored in its cache as follows:
\begin{align}
    \{W^{ij}_{d_k} \mid i, j \in [K] \setminus U_m, i \neq j\}, \quad \forall k \in U_m.
    \label{StepB}
\end{align}

\textbf{Transmission C:} The cache of the mirror contains two types of subfiles:
\begin{enumerate}
    \item \( W^{k(k+1)}_n - W^{kj}_n, \quad \forall n \in [N], \quad j \in S_k \), 
\( k \in U_m \),
    \item \( \sum_{n=1}^{N} W^{k(k+1)}_n, \quad \forall k \in U_m. \)
\end{enumerate}
After receiving the transmissions from the server to the mirrors, as described in \eqref{SM}, the mirror will acquire the subfile \( W^{k(k+1)}_{d_k} \) for each \( k \in U_m \). The mirror will then perform the following two transmissions:
\begin{equation} \label{StepC}
    \sum_{k \in U_m, \, k \text{ even}} W^{k(k+1)}_{d_k} \quad \text{and} \quad \sum_{k \in U_m, \, k \text{ odd}} W^{k(k+1)}_{d_k}.
\end{equation}


\textbf{Transmission D:} From the coded placement \( W^{k(k+1)}_n - W^{kj}_n \), as specified in \eqref{mp2}, the mirror can compute \( W^{kj}_{d_k}, \, \forall j \in S_k \). This process is detailed in Subsection~\ref{dec}, following Remark~\ref{remark1}. After obtaining these subfiles, the mirror then sends the following coded subfiles to the users:

\begin{equation}
    \sum_{k \in U_m} W^{kj}_{d_k}, \quad \forall j \in [K] \setminus (U_m \cup \{mK_2+1\}).
\label{stepd}
\end{equation}

Here, \( mK_2 + 1 \) denotes the first user associated with the \( (m+1) \)-th mirror. For the last mirror and its corresponding users, the first user of the next mirror is effectively the first user of the first mirror, creating a cyclic structure.

\textbf{Transmission E}: The mirror transmits the following.
\begin{equation}
    \sum_{k \in U_m \setminus \{s-1,s\}} W^{ks}_{d_k}, \quad \forall s \in (U_m \cup \{{mK_2+1}\}).
    \label{eq:ys}
\end{equation}
\begin{remark}\label{remark1}
For $K_2=2$, we have $U_m=\{2(m-1)+1, 2(m-1)+2\}$. Then Transmission E contains one redundant transmission, which corresponds to $s=2(m-1)+2$. 
\end{remark}

\subsubsection{Decoding (Proof of correctness)} \label{dec}
\paragraph{Transmission of files from server to users via mirrors}
In this subsection, we explain how each mirror, after receiving data from the server, acquires the necessary files for transmission to users during the delivery phase.

\textbf{For Transmission A}:
For \( k \in U_m \), the mirror transmits the data as specified in \eqref{SM}. For \( k \notin U_m \), using the cache content of the mirror as described in \eqref{cachmirr1}, the mirror can remove \( \sum_{k,s \in [K] \setminus U_m} \left( \frac{\alpha^s_k}{N^s_k} \right) W^{ks}_{d_s} \)  in order to decode the transmission given in \eqref{StepA2}.

\textbf{For Transmission B}: Since the cache of mirror \( m \) contains all the subfiles \( \{ W^{ij}_n \mid i, j \notin U_m, \, \forall n \in [N] \} \), it holds all the subfiles that need to be transmitted to users as specified in \eqref{StepB}.

\textbf{For Transmission C}: We know that the server sends the transmissions given in \eqref{StepA1} and for all \( k \in U_m \), the mirror holds in its cache \( \sum_{n=1}^{N} W^{k(k+1)}_n \) and \( W^{k(k+1)}_n - W^{kj}_n, \,\forall k \in U_m, \forall n \in [N], \, j \in S_k \).
Thus, the mirror can evaluate
\begin{align}
    \sum_{j \in S_k} \left( \frac{\alpha^j_k}{N^j_k} \right) \left( W^{k(k+1)}_{d_j} - W^{kj}_{d_j} \right), \quad \forall k \in U_m.
    \label{se1}
\end{align}
Now, combining \eqref{StepA1} and \eqref{se1}, we obtain:
\begin{align}
    \sum_{j \in S_k} \left( \frac{\alpha^j_k}{N^j_k} \right) W^{k(k+1)}_{d_j}, \quad \forall k \in U_m.
\label{se2}
\end{align}
So, this can be written as
\[
\sum_{j \in S_k, d_j \neq d_k} \left( \frac{\alpha_k^j}{N_k^j} \right) W_{d_j}^{k(k+1)} + \sum_{j \in S_k, d_j = d_k} \left( \frac{\alpha_k^j}{N_k^j} \right) W_{d_j}^{k(k+1)}.
\]
We know that \( \alpha_j^k = 1 \) when \( W_{d_j} \neq W_{d_k} \) and \( \alpha_j^k = -1 \) when \( W_{d_j} = W_{d_k} \). 
As each file is requested by at least one user, the above expression simplifies to
\[\begin{cases}
\sum_{n \in [N] \setminus d_k} W_n^{k(k+1)} & \text{if} \quad N_k^k = 0, \\
\sum_{n \in [N] \setminus d_k} W_n^{k(k+1)} - W_{d_k}^{k(k+1)} &  \text{if} \quad N_k^k > 0.
\end{cases}\]
The mirror can decode \( W_{d_k}^{k(k+1)} \) in both cases, as it has the following file in its cache
\[
\sum_{n=1}^{N} W_n^{k(k+1)} = \sum_{n \in [N] \setminus d_k} W_n^{k(k+1)} + W_{d_k}^{k(k+1)}, \quad \forall k \in U_m.
\]
Thus, it can proceed to send the transmissions as given in \eqref{StepC}.

\begin{remark}
For the scheme to operate effectively, the characteristic of the field must be at least \(K_1 K_2 - N + 2\), as \(N^j_k\) can have a maximum value of \(K_1 K_2 - N + 1\).
\end{remark}

\textbf{For Transmission D}: Now, the mirror holds \( W^{k(k+1)}_{d_k} \) for all \( k \in U_m \). Using its cached content \( W^{k(k+1)}_n - W^{kj}_n \), the mirror can compute \( W^{kj}_{d_k} \) for all \( j \in S_k \). Consequently, the mirrors are able to transmit the subfiles as described in \eqref{stepd}.

\textbf{For Transmission E}: We know that the mirrors now possess \( W^{kj}_{d_k} \) for all \( j \in S_k \) and \( k \in U_m \) as mentioned above. Therefore, they can proceed to send the transmissions outlined in \eqref{eq:ys}.
\paragraph{Successful file delivery to all users:} \label{ud}
In this subsection, we demonstrate that after receiving the transmissions from the mirrors, each user will obtain their requested file during the delivery phase.

Consider a user \( k \in U_m \) for some \( m \in [K_1] \). The demand of user \( k \) is \( W_{d_{k}} \), which it must recover using its cache content and the transmissions from its parent mirror \( m \). From the cache content of user \( k \), we know that it possesses \( C_k = \{ W^{ij}_n \mid \{i, j\} \cap U_m \neq \emptyset, i, j \in S_k, i \neq j, \forall n \in [N] \} \).
Now, the user needs the subfiles \( W^{ij}_{d_k} \) for \( i, j \notin U_m \), which the mirror sends directly to the user, as explained in Transmission B. The remaining subfiles for the user to retrieve are \( W^{kj}_{d_k} \) and \( W^{jk}_{d_k} \) for \( j \in S_k \).
Now, user \( k \) can decode \( W^{k(k+1)}_{d_k} \) from the transmissions in Transmission C using its cached content. This is possible because either \( k \) is even or odd, and \( W^{k'(k'+1)}_{d_k'} \in C_k \) for all \( k' \in U_m \setminus \{k,k+1\} \).

Then, from the transmissions listed in Transmission D and Transmission E, user \( k \) will be able to retrieve the remaining \( W^{kj}_{d_k} \) for \( j \in S_k \setminus \{k+1\} \), since it has \( W^{js}_{d_j} \) for all \( j \in S_k \) and \( s \in [K] \setminus U_m \).

The remaining subfiles to be retrieved are \( W^{jk}_{d_k} \) for \( j \in S_k \). These subfiles can be obtained from the transmissions sent in Transmission A from the mirror to the users, leveraging the user's cache content $C_k$.

\subsubsection{Rate}
\noindent \textbf{Server to mirrors:} \\
The total number of transmissions made in \eqref{SM} is $K$. Therefore, the rate is given by
\begin{equation}
R_1 = \frac{K}{2 \binom{K}{2}} = \frac{1}{K-1}.
\label{rate1}
\end{equation}
\noindent \textbf{Mirror to users:}
The total number of subfiles transmitted by the \( m^{\text{th}} \) mirror, where \( m \in [K_1] \), includes the following:
\begin{itemize}
    \item \( K \) coded subfiles in \textbf{Transmission A},
    \item \( K_2 \times 2 \binom{K - K_2}{2} \) uncoded subfiles in \textbf{Transmission B},
    \item \( 2 \) coded subfiles in \textbf{Transmission C},
    \item \( K - K_2 - 1 \) coded subfiles in \textbf{Transmission D}, and
    \item \( K_2 + 1 \) coded subfiles in \textbf{Transmission E}.
\end{itemize}
Therefore, the rate from the \( m^{\text{th}} \) mirror to its connected users is
\begin{align}
R_2 &= \frac{2 \left( K + 1 + K_2 \binom{K - K_2}{2} \right)}{K(K - 1)}.
\end{align}
\begin{remark}
As explained in Remark \ref{remark1}, for $K_2=2$, Transmission E contains one redundant transmission. Therefore, for $K_2=2$, $R_2 = \frac{2 \left( K + 1 + K_2 \binom{K - K_2}{2} \right)-1}{K(K - 1)}$.
\end{remark}

\subsection{The Second Scheme} \label{secondscheme}
The second scheme, inspired by the first, also operates for the case \( N \leq K_1 K_2 \). However, in this scheme, coded placement is performed directly in the users' caches instead of being stored in the mirrors' caches. This scheme operates for the memory pair \( (M_1, M_2) \), where
\begin{equation}
    M_1 = \frac{(K - K_2)(K - K_2 - 1)N}{K(K - 1)},
    \label{memory1second}
\end{equation}

\begin{equation}
    M_2 = \frac{N\left[ K(K - 2) - (K - K_2)(K - K_2 - 1) \right] + 1}{K(K - 1)}.
    \label{memory2second}
\end{equation}

\begin{table*}[ht]
\centering
\caption{Comparison of First Scheme with Other Schemes (\(R_1\), \(R_2\), \(\overline{R}\))} 
\resizebox{\textwidth}{!}{%
\begin{tabular}{|c c c|c c c|c c c|c c c|c c c|}
\hline
$N$ & $K_1$ & $K_2$ & \multicolumn{3}{c|}{First Scheme} & \multicolumn{3}{c|}{WWCY Scheme} & \multicolumn{3}{c|}{KNMD Scheme} & \multicolumn{3}{c|}{ZWXWLL Scheme} \\
 &  &  & $R_1$ & $R_2$ & $\overline{R}$ & $R_1$ & $R_2$ & $\overline{R}$ & $R_1$ & $R_2$ & $\overline{R}$ & $R_1$ & $R_2$ & $\overline{R}$ \\
\hline
3  & 1  & 3  & 0.5 & 1.16 & 1.66 & 0.47 & 1.41 & 1.88 & 0.11 & 1.65 & 1.76 & 0.47 & 1.41 & 1.74 \\ 
8  & 4  & 2  & 0.14 & 1.375 & 5.64 & 0.77 & 1.4 & 6.38 & 0.13 & 1.47 & 5.99 & 0.77 & 1.40 & 6.37 \\ 
10 & 5  & 2  & 0.11 & 1.47 & 7.5 & 0.79 & 1.50 & 8.28 & 0.10 & 1.56 & 7.90 & 0.79 & 1.5 & 8.29 \\ 
12 & 6  & 2  & 0.09 & 1.55 & 9.40 & 0.80 & 1.57 & 10.21 & 0.07 & 1.63 & 9.84 & 0.81 & 1.57 & 10.23 \\ 
14 & 7  & 2  & 0.07 & 1.60 & 11.34 & 0.81 & 1.62 & 12.16 & 0.06 & 1.67 & 11.80 & 0.81 & 1.62 & 12.15 \\ 
\hline  
\end{tabular}
}
\label{table first_scheme_comparison}
\end{table*}

\begin{table*}[ht]
\centering
\caption{Comparison of Second Scheme with Other Schemes (\(R_1\), \(R_2\), \(\overline{R}\))} 
\resizebox{\textwidth}{!}{%
\begin{tabular}{|c c c|c c c|c c c|c c c|c c c|}
\hline
$N$ & $K_1$ & $K_2$ & \multicolumn{3}{c|}{Second Scheme} & \multicolumn{3}{c|}{WWCY Scheme} & \multicolumn{3}{c|}{KNMD Scheme} & \multicolumn{3}{c|}{ZWXWLL Scheme} \\
 &  &  & $R_1$ & $R_2$ & $\overline{R}$ & $R_1$ & $R_2$ & $\overline{R}$ & $R_1$ & $R_2$ & $\overline{R}$ & $R_1$ & $R_2$ & $\overline{R}$ \\
\hline
3  & 1  & 3  & 0.5 & 0.5 & 1 & 0.73 & 0.73 & 1.46 & 0.74 & 0.74 & 1.48 & 0.74 & 0.74 & 1.48 \\ 
8  & 4  & 2  & 0.14 & 1.21 & 5 & 0.93 & 1.13 & 5.46 & 0.43 & 1.17 & 5.11 & 0.93 & 1.14 & 5.49 \\ 
10 & 5  & 2  & 0.11 & 1.35 & 6.86 & 0.99 & 1.27 & 7.34 & 0.34 & 1.31 & 6.89 & 0.99 & 1.27 & 7.34 \\ 
12 & 6  & 2  & 0.09 & 1.45 & 8.81 & 1.03 & 1.37 & 9.24 & 0.28 & 1.41 & 8.75 & 1.03 & 1.37 & 9.25 \\ 
14 & 7  & 2  & 0.07 & 1.52 & 10.71 & 1.10 & 1.44 & 11.18 & 0.24 & 1.49 & 10.67 & 1.10 & 1.44 & 11.18 \\ 
\hline
\end{tabular}
}
\label{table second_scheme_comparison}
\end{table*}

\subsubsection{Placement phase}
The placement for the $m^{\text{th}}$ mirror is given by
\begin{align}
   \{ W^{ij}_n \mid \ i, j \in [K] \setminus U_m, i \neq j, \forall n \in [N] \}.
   \label{cachmirr2}
\end{align}  
The following three types of subfiles are stored in the cache of user \( k \in [K] \).
\begin{itemize}
    \item 
    \begin{equation}
        \{ W^{ij}_n : \{i, j\} \cap U_m \neq \emptyset, \, i, j \in S_k, \, i \neq j, \, \forall n \in [N] \}.
        \label{up1}
    \end{equation}
    \item 
    \begin{align}
        W^{k(k+1)}_n - W^{kj}_n, \quad \forall n \in [N], \quad j \in S_k,
        \label{up2}
    \end{align}
    where \( S_k = [K] \setminus \{k\} \).
    \item 
    \begin{align}
        \sum_{n=1}^{N} W^{k(k+1)}_n.
        \label{up3}
    \end{align}
\end{itemize}

\subsubsection{Delivery phase}
\noindent \textbf{Server to mirrors:}
The transmission from the server to the mirrors follows the same procedure as described in the first scheme in \eqref{SM}.

\noindent \textbf{Mirror to users:} 

The mirror performs the following two transmissions to the users:

- The first transmission is equivalent to the transmission sent by the mirror to users in Transmission A of the first scheme.
- The second transmission corresponds to Transmission B of the mirror to users in the first scheme.

\subsubsection{Decoding}
\paragraph{Transmission of Files from Server to Users via Mirrors:}

As demonstrated in the first scheme, after receiving the server's transmissions and utilizing its cached content, each mirror should be capable of sending the appropriate transmissions to its connected users. Since the mirror's transmissions follow Transmission A and Transmission B from the first scheme, the decoding process also follows the same steps as outlined in Subsection \ref{dec} for these transmissions.

\paragraph{Successful file delivery to all users:}
From Transmissions A and B, any given user \( k \) can retrieve \( W^{jk}_{d_k} \) for \( j \in S_k \), as well as \( W^{ij}_{d_k} \) for \( i, j \notin U_m \), which mirrors the decoding process described in Paragraph~\ref{ud} for the first scheme. In the first scheme, the mirror stored the cached content \( \sum_{n=1}^{N} W^{k(k+1)}_n \) and \( W^{k(k+1)}_n - W^{kj}_n \), \( \forall k \in U_m, \forall n \in [N], \forall j \in S_k \), and used these to decode \( W_{d_k}^{k(k+1)} \), \( \forall k \in \mathcal{U}_m \). Here, the user itself stores \( W^{k(k+1)}_n - W^{kj}_n, \forall n \in [N], j \in S_k \), as well as \( \sum_{n=1}^{N} W^{k(k+1)}_n \). Therefore, using Transmissions A and B, it can decode \( W^{k(k+1)}_n \) in a similar manner to the mirror in the first scheme. Once \( W^{k(k+1)}_n \) is decoded, the user can combine it with its cached content \( W^{k(k+1)}_n - W^{kj}_n \) to obtain \( W^{kj}_{d_k} \) for \( j \in S_k \). The user also has \( \{ W^{ij}_n : \{i, j\} \cap U_m \neq \emptyset, i, j \in S_k, i \neq j, \forall n \in [N] \} \) in its cache. Now, combining all these recovered subfiles and its cache content, the user can reconstruct \( W_{d_k} \).

\subsubsection{Rate}
\noindent \textbf{Server to Mirrors:} \\
The total number of transmissions made from the server to the mirrors remains the same as in the first scheme, so \(R_1\) is the same as in \eqref{rate1}.

\noindent \textbf{Mirror to Users:} \\
Since the delivery from the mirror to users is the same as Transmissions A and B of the first scheme, the rate is calculated based on the transmissions given in \eqref{StepA1}, \eqref{StepA2}, and \eqref{StepB}. From \eqref{StepA1} and \eqref{StepA2}, a total of \(K\) transmissions are made from the \(m\)-th mirror to its attached users. Additionally, from \eqref{StepB}, a total of \(2 K_2 \binom{K-K_2}{2}\) transmissions are made. Therefore, the rate is 
\[
R_2 = \frac{K + 2 K_2 \binom{K-K_2}{2}}{K(K-1)}.
\]

\vspace{0.3cm}
\begin{example} \label{es, x1} 

We will explain one example on Scheme 1. Consider a system where the server has a library of \( N = 6 \) equally-sized files, denoted by \( \mathcal{W} = \{W_1, W_2, \ldots, W_6\} \). There are \( K_1 = 3 \) mirrors, with each mirror connected to \( K_2 = 2 \) users. Thus, the total number of users is \( K = K_1K_2 = 6 \).
The cache sizes are calculated as follows using \eqref{eq:M1} and \eqref{eq:M2}:
\begin{align*}
M_1 &= 4.067, \quad M_2 = 1.6.
\end{align*}

\textbf{Placement phase:}
The placement of the first mirror consists of the following files.
\[
\{ W^{ij}_n \mid i, j \notin \{1, 2\}, n \in [N], i, j \in [6] \},
\]
\[
W^{12}_n - W^{1k}_n, \quad \text{for} \ k \in \{3,4,5,6\},
\]
\[
W^{23}_n - W^{2k}_n, \quad \text{for} \ k \in \{1,4,5,6\},
\]
and
\[
\sum_{n=1}^{N} W^{12}_n, \quad \sum_{n=1}^{N} W^{23}_n.
\]
For the remaining two mirrors, the cache placement can be implemented using the same approach outlined above.

For the users attached to the first mirror, i.e., users 1 and 2, the placement is
\[
C_1 = \{ W^{ij}_n \mid i, j \in \{2, 3, 4, 5, 6\}, \, i \neq j, \, \forall n \in [6] \},
\]
\[
C_2 = \{ W^{ij}_n \mid i, j \in \{1, 3, 4, 5, 6\}, \, i \neq j, \, \forall n \in [6] \}.
\]

A similar placement can be established for users attached to other mirrors using \eqref{userplace}.

\textbf{Delivery Phase:}
In the delivery phase, let the demand vector be \( (1, 2, \ldots, 6) \). \\
\textbf{Server to mirrors}: The server transmits the following to the mirrors.
\[
\begin{aligned}
Y^1_{\overline{d}} &= W^{12}_{2} + W^{13}_{3} + W^{14}_{4} + W^{15}_{5} + W^{16}_{6}, \\
Y^2_{\overline{d}} &= W^{21}_{1} + W^{23}_{3} + W^{24}_{4} + W^{25}_{5} + W^{26}_{6}, \\
Y^3_{\overline{d}} &= W^{31}_{1} + W^{32}_{2} + W^{34}_{4} + W^{35}_{5} + W^{36}_{6}, \\
Y^4_{\overline{d}} &= W^{41}_{1} + W^{42}_{2} + W^{43}_{3} + W^{45}_{5} + W^{46}_{6}, \\
Y^5_{\overline{d}} &= W^{51}_{1} + W^{52}_{2} + W^{53}_{3} + W^{54}_{4} + W^{56}_{6}, \\
Y^6_{\overline{d}} &= W^{61}_{1} + W^{62}_{2} + W^{63}_{3} + W^{64}_{4} + W^{65}_{5}.
\end{aligned}
\]
\vspace{-0.3cm}


\[
\text{Therefore, the rate of the first layer is } R_1 = \frac{6}{6(6-1)} = 0.2.
\]
\textbf{Mirror to users}: The first mirror transmits the following transmissions to its attached users using \eqref{StepA1} and \eqref{StepA2}.
\vspace{-0.3cm}

\begin{figure*}[!t]
    \centering
   \includegraphics[ width=0.69\textwidth]{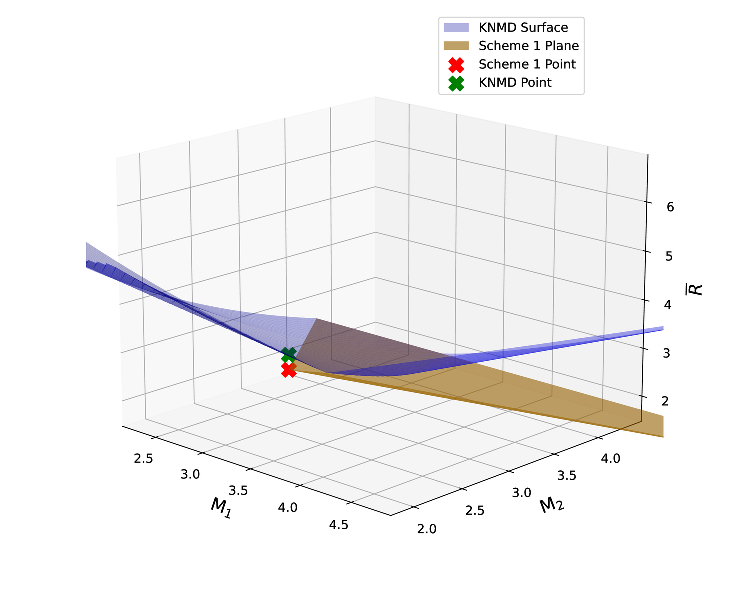}
    \caption{Comparison of composite rate \( \overline{R} \) for the ($3,2$;\emph {$M_{1}$},\emph{$M_{2}$};6) hierarchical caching system in Example \ref{es, x1}.}
\label{fig1}
\end{figure*}

\begin{figure*}[!t]
    \centering
   \includegraphics[ width=0.69\textwidth]{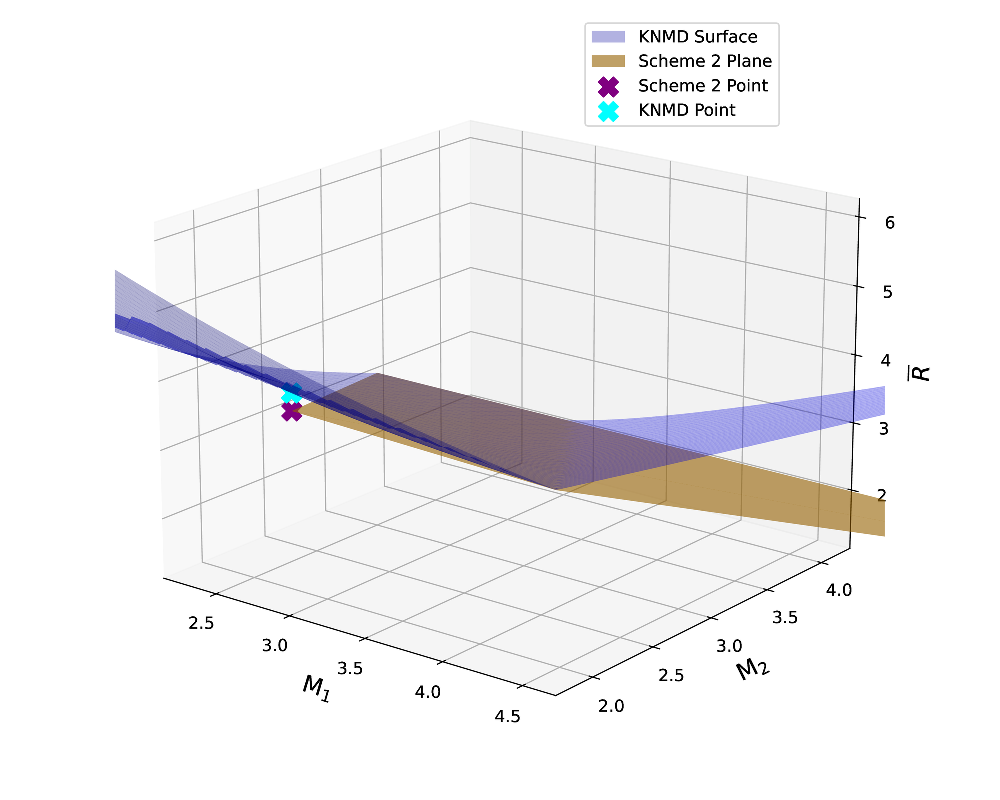}
  \caption{Comparison of Scheme 2 with KNMD scheme \cite{karamchandani2016hierarchical} for the ($3,2$;\emph {$M_{1}$},\emph{$M_{2}$};6) hierarchical caching system.}

    \label{plot}
\end{figure*} 

\[
\begin{aligned}
    & W^{12}_{2} + W^{13}_{3} + W^{14}_{4} + W^{15}_{5} + W^{16}_{6}, \\
    & W^{21}_{1} + W^{23}_{3} + W^{24}_{4} + W^{25}_{5} + W^{26}_{6}, \\
    & W^{31}_{1} + W^{32}_{2}, \quad W^{41}_{1} + W^{42}_{2}, \\
    & W^{51}_{1} + W^{52}_{2}, \quad W^{61}_{1} + W^{62}_{2}.
\end{aligned}
\]
\vspace{0.4cm}
The second type of transmissions, as outlined in \eqref{StepB}, are
\vspace{-0.3cm}
 \begin{align*}
    W^{34}_1, W^{35}_1, W^{36}_1, W^{45}_1, W^{46}_1, W^{56}_1, & \\
    W^{43}_1, W^{53}_1, W^{63}_1, W^{54}_1, W^{64}_1, W^{65}_1, & \\
    W^{34}_2, W^{35}_2, W^{36}_2, W^{45}_2, W^{46}_2, W^{56}_2, & \\
    W^{43}_2, W^{53}_2, W^{63}_2, W^{54}_2, W^{64}_2, W^{65}_2.
\end{align*}

Next, consider the third type of transmission: The mirror sends the subfiles \( W^{12}_1 \) and \( W^{23}_2 \).
For the fourth type of transmissions, using \eqref{stepd}, the mirror sends \( W^{14}_1 + W^{24}_2 \), \( W^{15}_1 + W^{25}_2 \), and \( W^{16}_1 + W^{26}_2 \).
Finally, for the fifth type, as per \eqref{eq:ys}, the mirror transmits \( W^{13}_1 \) and \( W^{21}_2 \).

Therefore, the rate for the second layer is

\[
R_2 = \frac{6 + 24 + 2 + 3 + 2}{6(6-1)} = \frac{37}{30}=1.23.
\]

Similarly, transmissions from the other mirrors to their respective users also yield \( R_2 = 1.23 \). Thus, the composite rate is given by 
\[
\overline{R}=R_1 + K_1 R_2 = 0.2 + 3 \times \frac{37}{30}= 3.9.
\]
\vspace{-0.2cm}
\\
For the same setup with \(K_1 = 3\), \(K_2 = 2\), the second scheme yields \(M_1 = 2.4\), \(M_2 = 2.43\), \(R_1 = 0.2\), \(R_2 = 1.0\), and \(\overline{R} = R_1 + K_1 R_2 = 3.2\).

We compare both of our schemes for the setup \( (K_1 = 3, K_2 = 2, N = 6) \) with the KNMD scheme from \cite{karamchandani2016hierarchical} in Figures \ref{fig1} and \ref{plot}. While the KNMD scheme in \cite{karamchandani2016hierarchical} applies to all memory points \( (M_1, M_2) \), our schemes are evaluated at specific pairs of memory points \( (M_1, M_2) \). We consider the trivial points \( (M_1 = 0, M_2 = N, R_1 = R_2 = 0) \), \( (M_1 = N, M_2 = 0, R_1 = 0, R_2 = K_2) \), and \( (M_1 = N, M_2 = N, R_1 = R_2 = 0) \), and apply the memory sharing approach for hierarchical two-layer coded caching systems as outlined in \cite{PRRarXiv} to define the planes for our schemes in Figures \ref{fig1} and \ref{plot}.


Figures \ref{fig1} and \ref{plot} focus solely on comparing our schemes with the KNMD scheme \cite{karamchandani2016hierarchical}, as it outperforms the schemes in \cite{8481555,wang2019reduce}. A detailed comparison with all these schemes is provided in Tables \ref{table first_scheme_comparison} and \ref{table second_scheme_comparison}. Including multiple schemes in the 3D figures for the composite rate \( \overline{R} \) as a function of \( M_1 \) and \( M_2 \) would result in visual clutter due to overlapping planes, so the figures are limited to the KNMD scheme for clarity.

From Figure \ref{fig1}, at the memory pair \( (M_1 = 4.06, M_2 = 1.6) \), the KNMD scheme achieves a composite rate of \( \overline{R} = 4.20 \), while our first scheme gives \( \overline{R} = 3.90 \). Similarly, Figure \ref{plot} shows that at the memory pair \( (M_1 = 2.4, M_2 = 2.43) \), the KNMD scheme provides \( \overline{R} = 3.47 \), whereas our second scheme achieves \( \overline{R} = 3.2 \).

\end{example}

Table \ref{table first_scheme_comparison} compares our proposed first scheme with the schemes in \cite{karamchandani2016hierarchical,8481555,wang2019reduce} for different values of \( K_1 \), \( K_2 \), and \( N \). Similarly, Table \ref{table second_scheme_comparison} presents a comparison of our second scheme with these schemes. Both tables provide the corresponding \( (R_1, R_2) \) pairs for each scheme, for the memory pairs \( (M_1, M_2) \) achieved by our two schemes, which lie in the high memory regime (i.e., when both \( M_1 \) and \( M_2 \) are large), and the values of the composite rate \( \overline{R} \) for these schemes and our schemes at these memory points.

From Table \ref{table first_scheme_comparison}, we observe that our first scheme achieves a reduced composite rate \( \overline{R} \) for smaller values of \( K_2 \), irrespective of the value of \( K_1 \). In contrast, Table \ref{table second_scheme_comparison} highlights the performance of our second scheme compared to other schemes. It can be seen that for larger values of \( K_1 \), the composite rate \( \overline{R} \) of the KNMD scheme is lower than that of our second scheme. This is because the \( R_2 \) value for KNMD is smaller. However, our second scheme achieves a significantly better \( R_1 \) compared to KNMD. Since \( \overline{R} = R_1 + K_1 R_2 \), the larger \( R_2 \) value for our scheme slightly increases the composite rate \( \overline{R} \) for higher \( K_1 \).

Based on these observations, we can conclude the following:

\begin{itemize}
 \item If the application prioritizes a lower transmission load from the server (\( R_1 \)), Scheme 2 is preferable, as it significantly outperforms KNMD in terms of \( R_1 \). The \( R_1 \) values of our second scheme are consistently lower than those of KNMD.

\item If the focus is on achieving a lower composite rate \( \overline{R} \), Scheme 1 is a better choice. This scheme achieves a reduced \( \overline{R} \), particularly for larger \( K_1 \), while the difference in \( R_1 \) between our scheme and KNMD is smaller in this case.

\end{itemize}

\section{Conclusion}
In this study, we address the two-layer hierarchical coded caching problem by introducing a coded placement scheme. Our schemes significantly lower transmission rates compared to existing approaches, particularly in high memory regimes defined by large values of \( M_1 \) and \( M_2 \).
We demonstrate this improvement by comparing results using a 3D figure. Future work includes identifying additional memory points or possibly developing a generalization that works across a wider range of memory configurations.

\section*{Acknowledgment}
This work was supported partly by the Science and Engineering Research Board (SERB) of the Department of Science and Technology (DST), Government of India, through the J.C. Bose National Fellowship awarded to B. Sundar Rajan, by the Ministry of Human Resource Development (MHRD), Government of India, through the Prime Minister’s Research Fellowship (PMRF) awarded to Rajlaxmi Pandey, and by the Wallenberg AI, Autonomous Systems and Software Program through a joint project grant awarded to Aalto University and Chalmers University of Technology (PIs A. Graell i Amat and C. Hollanti), which supported Charul Rajput’s research.

\section*{Appendix: Derivation of Memory Expressions \(M_1\) and \(M_2\)}

The memory allocation expressions \(M_1\) and \(M_2\) are derived based on the placement strategies described in the manuscript. We provide derivations separately for Scheme 1 and Scheme 2.

\subsection*{\textbf{Scheme 1}}
\subsubsection*{Mirror Memory \(M_1\) (Scheme 1)}
The memory allocation \(M_1\) for each mirror in Scheme 1 is determined by three types of placement:

\begin{itemize}
    \item The first type of placement, described in \eqref{cachmirr1}, stores \(2 \cdot \binom{K-K_2}{2} \cdot N\) subfiles in each mirror, where each subfile has a size of \(\frac{1}{K(K-1)}\).
    
    \item The second type of placement, described in \eqref{mp2}, stores \(K_2 \cdot N \cdot (K-2)\) subfiles in each mirror, where each subfile also has a size of \(\frac{1}{K(K-1)}\).
    
    \item The third type of placement, described in \eqref{mp3}, stores \(K_2\) subfiles, each of size \(\frac{1}{K(K-1)}\).
\end{itemize}

By summing the contributions from these three types of placement and simplifying, the total memory required at each mirror is:
\begin{equation}
M_1 = \frac{(K-K_2)(K-K_2-1)N + K_2(K-2)N + K_2}{K(K-1)},
\label{eq:M1_appendix_scheme1}
\end{equation}
as stated in \eqref{eq:M1}.

\subsubsection*{User Memory \(M_2\) (Scheme 1)}
The memory allocation \(M_2\) for each user in Scheme 1 is derived from the placement strategy in \eqref{userplace}. Each user stores \(2N \cdot \left[\binom{K-1}{2} - \binom{K-K_2}{2}\right]\) subfiles, where each subfile has a size of \(\frac{1}{K(K-1)}\). 

Hence, the total memory required at each user is:

\begin{equation}
M_2 = \frac{[(K-1)(K-2) - (K-K_2)(K-K_2-1)]N}{K(K-1)},
\label{eq:M2_appendix_scheme1}
\end{equation}
as stated in \eqref{eq:M2}.

\subsection*{\textbf{Scheme 2}}

\subsubsection*{Mirror Memory \(M_1\) (Scheme 2)}

In Scheme 2, the memory allocation \(M_1\) for each mirror is based on the placement strategy in \eqref{cachmirr2}. Each mirror stores \(2 \cdot \binom{K-K_2}{2} \cdot N\) subfiles, where each subfile has size \(\frac{1}{K(K-1)}\). 

The total memory per mirror is:
\[
M_1 = 2 \cdot \binom{K-K_2}{2} \cdot N \cdot \frac{1}{K(K-1)}.
\]

Using \(\binom{K-K_2}{2} = \frac{(K-K_2)(K-K_2-1)}{2}\), this simplifies to:

\[
M_1 = \frac{(K-K_2)(K-K_2-1)N}{K(K-1)}.
\]

Thus, the memory required at each mirror is:
\begin{equation}
M_1 = \frac{(K-K_2)(K-K_2-1)N}{K(K-1)},
\label{eq:M1_appendix_scheme2}
\end{equation}
as stated in \eqref{memory1second}.

\subsubsection*{User Memory \(M_2\) (Scheme 2)}

The memory allocation \(M_2\) for each user in Scheme 2 is derived based on three types of placement:

\begin{itemize}
    \item The first type of placement, as described in \eqref{up1}, stores 
    
    \[
    2N \left[\binom{K-1}{2} - \binom{K-K_2}{2}\right]
    \]
    subfiles.

    \item The second type of placement, described in \eqref{up2}, stores \(N(K-2)\) subfiles.

    \item The third type of placement, described in \eqref{up3}, stores a single subfile.
\end{itemize}

The size of each subfile is \(\frac{1}{K(K-1)}\). Summing the contributions, the total memory per user is:

\[
M_2 = \frac{2N \left[\binom{K-1}{2} - \binom{K-K_2}{2}\right] + N(K-2) + 1}{K(K-1)}.
\]

Simplifying, this becomes:

\begin{equation}
M_2 = \frac{N\left[K(K-2) - (K-K_2)(K-K_2-1)\right] + 1}{K(K-1)},
\label{eq:M2_appendix_scheme2}
\end{equation}
as stated in \eqref{memory2second}.


\end{document}